**Ferroelectricity and strain effects in orthorhombic $YMnO_3$ thin films**

J. Fontcuberta, I. Fina, L. Fàbrega, F. Sánchez
Institut de Ciència de Materials de Barcelona (ICMAB-CSIC), Campus UAB s/n, 08193 Bellaterra, Spain

V. Skumryev
Institució Catalana de Recerca i Estudis Avançats (ICREA) and Departament de Física, Universitat Autònoma de Barcelona, 08193 Bellaterra, Spain

X. Martí
Department of Condensed Matter Physics, Faculty of Mathematics and Physics, Charles University, Ke Karlovu 5, 121 16 Praha 2, The Czech Republic.

ABSTRACT

We report on the dielectric properties of epitaxial [001] and [100]-textured thin films of antiferromagnetic orthorhombic $YMnO_3$ and their variation under compressive epitaxial strain. It is found that weakly strained $YMnO_3$ films are ferroelectric with a polarization along *c*-axis switchable by 90º by an external magnetic field. When reducing film thickness and increasing epitaxial strain, the ferroelectric order is progressively suppressed. Analysis of structural, dielectric and magnetic data indicates that suppression of ferroelectricity when reducing thickness is accompanied by an enlarged ratio *a*/*b* of the in-plane cell parameters of the orthorhombic structure and the appearance of a net magnetization. All results can be well described by considering the multiferroic phase diagram of orthorhombic manganites.

## Introduction

The discovery of ferroelectricity induced by magnetic order in orthorhombic $RMnO_3$ (where R is a lanthanide) perovskites [1, 2] has triggered intensive research aiming to understand its microscopic origin and eventually to exploit the coupled magnetic and ferroelectric responses in practical devices. A fundamental parameter to rationalize their complex magnetic phase diagram is the size of the lanthanide cation (R) and subsequently the orthorhombic distortion, which is commonly parametrized by the Mn-O-Mn bonding angle ($\theta$) although other parameters, such as the *a/b* ratio where *(a,b,c*) are the cell parameters (the Pbnm setting is used) can also be used. It is known that for large R (i.e. $LaMnO_3$) the bond angle $\theta$ and the *a/b* (<1) ratio are the largest, indicating a small distortion, and the magnetic structure is a collinear A-type antiferromagnet (the first-neighbour ($J_1$) interactions within the *ab* plane are ferromagnetic but antiferromagnetic along the *c*-axis). As the size of the lanthanide is reduced, the rotation of the $MnO_6$ octahedra leads to the closing of the bonding angle $\theta$, a reduction of *a/b* and the magnetic interactions among next-neighbours within the *ab* plane ($J_2$) (antiferromagnetic in nature) get relatively more important; consequently there is an increasing competition of magnetic interactions ($J_2/J_1$ enlarges) producing complex non-collinear, modulated magnetic orders. In $TbMnO_3$ for instance, when lowering the temperature, the magnetic structure first orders in a collinear sinusoidal spin arrangement below the Néel temperature ($T_N$) which, when further cooling, transforms into a *bc*-cycloidal. In agreement with current understandings [3-5], a ferroelectric polarization develops along *c*-axis [1, 2] as $\mathbf{P}_c = A\ \mathbf{r}_{ij} \times (\mathbf{S}_i \times \mathbf{S}_j)$ where ($\mathbf{S}_i$, $\mathbf{S}_j$) are neighbouring Mn ionic spins and $\mathbf{r}_{ij}$ is the unit vector connecting them. When the size of R is further reduced $\theta$ and *a*/*b* lower further, the in-plane magnetic structure changes to a collinear E-type order, with characteristic ferromagnetic zig-zag chains along the *a*-axis and antiferromagnetic coupling along *b*-axis. Ferroelectricity [6, 7] in the collinear E-type antiferromagnets is driven by the asymmetric magnetostriction along *b* and *a*-axis which breaks the center of symmetry of the lattice producing a characteristic finite polarization along *a*-xis ($\mathbf{P}_a \sim \mathbf{S}_i \cdot \mathbf{S}_j$) [8-10].

Therefore *bc*-cycloidal and E-type antiferromagnetic ferroelectrics have distinctive orientation of the induced ferroelectricity: $P_c$ and $P_a$ respectively. In addition, due to the sensitivity of orientation of the cycloidal plane to external magnetic fields, magnetic-field induced switching from *bc* to *ac*-cycloids, accompanied by subsequent switching of polarization from $P_c$ to $P_a$ (90º switching) have been reported in single crystals [1, 2] and was more recently observed in thin films [11].

When $RMnO_3$ oxides are grown on single crystalline substrates, epitaxial strain is expected to induce changes on bond angles and bond lengths thus leading to substantial changes on their properties. Whereas strain effects have been long documented in hole-doped manganites (i.e $R_{1-x}B_xMnO_3$, where R is a trivalent rare earth and A is a divalent alkaline earth, see for instance [12] and references therein, systematic studies in multiferroic $RMnO_3$ thin films are only emerging [13-15]. For instance, strain effects in $YMnO_3$ (YMO) have been recently reported and dramatically distinct responses of relaxed and strained films have been observed. Relatively thick (~100 nm) YMO films grown on $SrTiO_3$ substrates display clear antiferromagnetic character: a well defined cusp in the magnetic susceptibility ($\chi$) at the Néel temperature ($T_N \approx 40$ K), the expected magnetic anisotropy of the susceptibility with respect to the antiferromagnetic axis (*b*-axis) at $T<T_N$, and reversible zero-field cooling and field-cooling magnetization [14]. Ferroelectric polarization measurements on *a*-textured YMO films indicate that, under a magnetic field applied along *b*-axis, the polarizacion increases rapidly up to about 80 $nC/cm^2$. In parallel, measurements on *c*-textured YMO films indicate the existence of a finite polarization $P_c$ that lowers under magnetic field [11]. These data taken together indicate that in thin film form, YMO has a cycloidal magnetic structure. This finding contrasts to what had been reported for bulk YMO, where neutron diffraction experiments were interpreted in terms of an E-type structure [16]. However, pyrocurrent measurements in polycrystalline samples provided polarization values of about 100 $nC/cm^2$ [6, 7] much in line with what is expected for cycloidal than for E-type, where P values are expected to be of about two orders of magnitude larger [8-10].

Strain effects on the properties of these films have most evident hallmarks in their magnetic properties [14]. It has been found that when reducing the film thickness, a clear irreversibility appears in the magnetization *vs* temperature curves M(T) and magnetization *vs* magnetic field loops M(H). This irreversibility signals the occurrence of a net magnetic moment, even at remanence, which has been taken as evidence of a canted state in the magnetic ordering and of a non-collinear magnetic order that develops and increases when reducing the film thickness. Structural analysis of these films indicated that the cell parameters of the films are in-plane anisotropically strained and the unit cell volume is compressed, both gradually relaxing as the film thickness grow.
Measurements of the dielectric permittivity of these films under magnetic field have revealed a substantial magnetic-field dependence of this parameter and thus a clear magnetoelectric coupling. Interestingly enough, it has been reported that the magnetoelectric coupling depends on film thickness. More precisely, it has been shown that it reduces for the most strained, thinner films and thus for films displaying the larger magnetization [13].

It is the purpose of this manuscript to provide a unified view of all these results. We report here structural, dielectric and magnetic data on epitaxial YMnO thin films of [100] and [001] textures of different thickness. These sets of films allows us to explore the anisotropy of the dielectric response and strain-related effects. On the basis of the present data, the recently observed cycloidal order in thick YMO films [11] and reported gradual reduction of magnetoelectric coupling under strain [13], we will argue that compressive epitaxial strain acting on YMO produces a shift from cycloidal to collinear magnetic structure, most likely A-type, with subsequent loss of polarization and the appearance of a canted magnetic structure.

**Experimental**

Epitaxial films of $YMnO_3$ (YMO) were grown on (001) or (110) Nb:$SrTiO_3$ (STO) substrates. Further information on sample preparation and structural characterization methods for these films can be found elsewhere [17]. In short, YMO grows respectively [001] (*c*-axis) and [100] (*a*-axis) textured on these substrates (see below). X-ray diffraction phi-scans (not shown here) [17] indicated that *c*-textured films contain two in-plane 90º-rotated, families of crystallites whereas the *a*-textured are single domain. Films of different thicknesses *t* (20 nm < t < 140 nm) have been grown.

Dielectric measurements on [100] and [001] films were performed using a planar capacitor configuration with two Pt-top contacts (0.15 $mm^2$) and the conducting Nb:STO as bottom electrode, allowing to extract the dielectric permittivity along the *a*-axis ($\varepsilon_a$) and along the *c*-axis ($\varepsilon_c$), respectively. In these experiments an excitation voltage of 300 mV at 100 kHz was used. Current versus electric field loops (I-E) were recorded at 10 kHz using the Positive-Up-Negative-Down (PUND) technique (TF Analyser 2000 from AixACCT Co) [18]. Dielectric measurements at low temperature and under magnetic field have been carried out using a PPMS from Quantum Design [19]. Temperature dependent runs were performed using a sweep rate of 5 K/min. Magnetic measurements were performed using a SQUID magnetometer with the magnetic field applied parallel to the sample surface.

**Results**

X-ray diffraction pole figures collected at the YMO(111) scattering angle (Figs. 1(a) and (e)) reveal that the films grown on STO(110) and STO(001) are [100]- (*a*-axis) and [001]-textured (*c*-axis), respectively. The visible spots corresponding to the main texture are labelled (large white arrow) in Figs. 1. In addition to the main texture, films display minor presence of [010] and [110] oriented material for *a*- and *c*-oriented films, respectively; the corresponding spots in the pole figures are labelled by a short black arrow.

The lattice parameters of the films have been investigated by X-ray diffraction. Following the procedures described in Ref. [17], the unit cell dimensions for *a*-oriented films of various thickness have been obtained from phi-scans (not shown) and reciprocal space maps around the (311) reflections (Figs. 1(b-d)). Bulk positions are signaled by 'X' symbols in each panel. The cell parameters (*a*, *b*, *c*) are collected in Table I. As shown by data in Table I, the unit cell along the out-of-plane direction (*a*) is identical to the bulk value [20] thus indicating fully relaxation irrespectively of film thickness. On the other hand the in-plane cell parameters are anisotropically strained: the *b* axis is remarkably shorter than its bulk counterpart and close, for the thinnest film, to the substrate in-plane parameter ($\sqrt{2}\, a_{STO}$ = 5.52 Å where $a_{STO}$ is the cell parameter of $SrTiO_3$) and the *c* axis is longer and both relax approaching bulk value when increasing film thickness. In summary, the epitaxial strain is anisotropic being tensile and compressive along [001] and [010] in-plane directions, respectively, and negligible along [100] out-of-plane direction. Correspondingly, the unit cell volume ($V_{uc}$) is compressed under the stress imposed by the substrate and gradually tend to the bulk one ($V_{uc}$ = 226.4 $Å^3$) as thickness increases.

Similarly, in Figs. 1 (f-h) we show the reciprocal spaces maps of the (208) (right) and (028) (left) reflections of the *c*-textured YMO films of various thicknesses. Bulk positions are signaled by 'X' symbols in each panel. The extracted cell parameters are also collected in Table I. While the (208) reflections appear at the bulk position, thus indicating that the cell along *a* is fully relaxed, the (028) reflections are shifted towards higher Q-values indicating smaller *b* parameters than in bulk YMO; therefore the unit cells are compressively strained along *b*. Regarding the out-of-plane direction, an expansion of the *c* parameter is observed. Therefore the observed trend in *c*-textured films is identical to that described above for *a*-textured films: the unit cell is fully relaxed along *a* for all films whereas the *b* axis contracts and *c*-axis expands as thickness is reduced. Notice that the thickest films of *a*- and *c*-texture (105 nm and 137 nm, respectively) having similar $V_{uc}$ values (≈ 224.0 $Å^3$) display the persistence of a residual strain (for bulk YMO. $V_{uc}$ = 226.4 $Å^3$).

We turn now to the dielectric properties, measured using the contact configuration displayed in Figs. 2(a,d). For clarity we show in Figs. 2 (b,e) sketches indicating the crystallographic axis for samples of both textures. The existence of two in-plane crystallites in *c*-textured films is indicated. In Figs. 2c and 2f we show the temperature dependence of the $\Delta\varepsilon_a$ and $\Delta\varepsilon_c$ for *a*- and *c*-textured films of distinct thicknesses. Here in order to better visualize the low-temperature evolution of the permittivity, the relative changes $\Delta\varepsilon_{a,c}(T) = [\varepsilon_{a,c}(T) - \varepsilon_{a,c}(60K)]/\varepsilon_{a,c}(60\ K)$ are displayed. In both cases, the permittivity increases below about 40 K, coinciding with the Néel temperature, and develops a broad but well defined peak at around 20 K. The occurrence of this anomaly in the permittivity is a signature of the appearance of ferroelectricity. This has been further confirmed by measuring the electric field dependence of the dielectric permittivity, its temperature dependence and the I-V characteristics. In Figs. 3(a-f) (left axis) we show the $\varepsilon_a(E)$ and $\varepsilon_c(E)$ measured at 5K, for the *a*- and *c*-textured films respectively. The corresponding losses are indicated in the right axis. It is clear that all $\varepsilon_a(E)$ and $\varepsilon_c(E)$ display the butterfly-like shape typical of ferroelectrics, well visible even for the thinner films. ε(E) loops become progressively asymmetric when reducing film thickness; this effect can be attributed to the asymmetry of the electrodes and to the small ac-voltage applied (up to 300 mV) in these measurements. For the thicker films of each texture (106 nm and 137 nm) we have recorded the temperature dependence of $\varepsilon_a(E)$ and $\varepsilon_c(E)$, and used these data to evaluate the integrated area (Q) within each $\varepsilon_{a,c}(E)$ loop, which can be taken as a measure of the polarization and its temperature dependence. In Fig. 4 we show the extracted values of Q for both films; these data clearly reveal that a polarization develops below 25 K. Further confirmation of the ferroelectric character of all these samples has been obtained from the I-V characteristics measured at 5 K. In Figs. 5 the I-V curves collected for the distinct pulses of the PUND procedure are included. For those corresponding the $t$ = 106 nm and 137 nm *a*-textured and *c*-textured films (Figs. 5(a) and (d), respectively), the polarization switching current peak is well evident; the corresponding integration leads to polarization values of $P_a$ = 80 nC/cm$^2$ and $P_c$ = 90 nC/cm$^2$ [11]. Obviously, the signatures of switching current become weaker when reducing film thickness (Figs. 5(c) and (f)); in this case, integration does not provide reliable numerical values of P, and therefore those are not given.

Now, we focus our attention on the core of this manuscript, that is the magnetic field (H) dependence of the permittivity ($\varepsilon_a$, $\varepsilon_c$) and its anisotropy. In Figs. 6(a-c) the field-dependent permittivity ($\Delta\varepsilon(H, 5K) = [\varepsilon(H)-\varepsilon(H=0)]/\varepsilon(H=0)$) measured at 5 K on the *a*-textured films of various thicknesses is shown for different orientations of H. In Fig. 6(a), where data corresponding to the $t$ = 106 nm film are shown, it can be appreciated that $\Delta\varepsilon_a(H)$ (and $\varepsilon_a(H)$) displays a well pronounced peak at about 4.6 T for H//*b*, which is clearly absent for H//*a* and H//*c*. This peak appears concomitantly with the reported field-induced polarization along $P_a$ occurring at this field and attributed to a switching of polarization from *c*-axis to *a*-axis [11] in close analogy to what had been reported in $TbMnO_3$ single crystals [1]. The lack of any anomaly in $\varepsilon_a(H)$ when H is applied in a direction other than *b*, confirms the assignment of a *bc*-cycloidal to the magnetic structure of the thicker (less strained) film. For H > 4T the permittivity decreases and displays a smooth field dependence. As film thickness is reduced (Fig. 6(b)) and strain increases, the dielectric peak is still visible at a higher magnetic field (4.8 T for $t$ = 50 nm film) although the value of the permittivity $\varepsilon_a$(H) at its position is largely reduced (notice the difference in the y-axis scales). For the thinnest film (Fig. 6(c)) only traces of switching are visible in the $\varepsilon_a$(H//b) around 5.3 T. It can also be appreciated that the peak in $\Delta\varepsilon_a$(H) is hysteretic; a detailed analysis of this behaviour is beyond the objectives of the present work. The corresponding $\Delta\varepsilon_c$(H) data for the *c*-textured films of various thicknesses are shown in Fig. 6(d-f). In sharp contrast to $\Delta\varepsilon_a$(H), $\Delta\varepsilon_c$(H) displays a weaker dependence, almost monotonic, on magnetic field. Measurements have also been performed with the magnetic field applied along the *c*-axis and *ab*-plane. No significant dependence of $\Delta\varepsilon_c$(H) on the orientation of H has been observed. In all cases, when decreasing thickness, the variation of $\Delta\varepsilon_c$(H) weakens. It is worth to remark that well defined $P_c$(E) loops have been observed for the thickest ($t$ = 137 nm) film which is fully compatible with the existence of *bc*-cycloids generating $P_c \neq 0$. The application of H//*ab* produced a reduction of $P_c$ as expected from the switching from *bc* to *ac*-cycloidal plane [1]. Correspondingly, signatures of switching should be visible in $\Delta\varepsilon_c$(H) on *c*-textured films. Indeed, a close inspection of the $\Delta\varepsilon_c$(H) data (see insets in Figs 6(d-f)) evidences the associated signature around 4-5 T.

Having established that the *a*-textured films display a field-induced switching of P that gradually disappears when reducing film thickness it is of relevance to analyze in detail the magnetoelectric response of the strained films. In Figs. 7(a) and (b) we show the relative change of permittivity ($\Delta\varepsilon_a$(H) and $\Delta\varepsilon_c$(H)) *vs* $H^2$. Data in Fig. 7(b) clearly indicate that the change in $\Delta\varepsilon_c$(H) permittivity is quadratic in magnetic field. A similar trend can be inferred from the high-field region of data $\Delta\varepsilon_a$(H). This quadratic dependence is more apparent when the film thickness is reduced and traces of the field-induced switching disappear.

Magnetic characterization has been done by measuring the temperature dependence of the magnetization of all films upon warming up after both zero-field-cooling (ZFC) and field cooling (FC) processes using 50 mT field (H//*a*). Data for the *a*-textured films of distinct thicknesses are shown in Fig. 8(a). The corresponding data for *c*-textured films have been reported elsewhere [14]. Data in Fig. 8(a) show a cusp at the Néel temperature. Importantly, the FC magnetization increases when film thickness is reduced and, in close agreement to what has been reported for *c*-textured films, a clear splitting of the ZFC-FC curves hysteresis develops, signalling the occurrence of a net magnetization at $T < T_N$. These features were attributed to the existence of a strain-induced canting that becomes more relevant as the film thickness is reduced [14]. The presence of the hysteresis in the ZFC-FC curves and the concomitant presence of a net magnetization are also reflected in the existence of a magnetic remanence in the corresponding magnetization loops M(H). Indeed, we have used the M(H) loops measured at 25 K (H//*c*) to determine the remanent magnetization ($M_r$), which naturally increases when reducing film thickness. The magnetization M(25K, 50 mT) and the remanence $M_r$(5K) of all here reported *a*-textured films are depicted in Fig. 9 (solid and open circles respectively) as a function of the measured *a/b* ratio. Data in this plot reveal the enhanced net magnetic moment as film thickness is reduced and the corresponding *a/b* ratio increases. We have also included in Fig. 9 the M(25 K, 50 mT) magnetization values (not shown here) of the *c*-textured films of different thickness (solid triangles) as well as data earlier reported on similar *c*-textured films [13,14]. It is clear that all data show the same trend, that is the net magnetic moment increases when increasing strain and the *a/b* ratio.

**Discussion.**

The data described above can be summarized as follows: a) for both *c*-textured and *a*-textured films, when reducing film thickness strain develops across the films and the unit cell volume is compressed; b) $\varepsilon_a$(T,H) in *a*-textured films evidences the ferroelectric character and field-induced 90º switching of polarization. These are most visible in the thicker films; they gradually disappear as film thickness is reduced and, according to structural data, compressive strain is reinforced. We interpreted this as a signature of the existence of a *bc*-cycloid in the thicker *a*-textured films; c) $\varepsilon_c$(T,H) data shows a moderate reduction under field and only tiny signatures of switching; d) in all cases, the permittivity shows a weaker (quadratic) field dependence when reducing thickness. We interpreted this complete set of results as signatures of destabilization of the cycloidal order towards some basically collinear magnetic order under strain. On the other hand, previous reports have shown that under strain, *c*-textured films display a weak net magnetization that has been attributed to a strain-induced canted magnetic structure. More precisely, it has been proposed that the fan of canted spins spans up to about $1^o$ away from b-axis [14]. The present data for *a*-textured films follow the same trend.

Discussion of these results should start by considering the multiferroic phase diagram of $RMnO_3$ perovskites. As mentioned in the introduction, the multiferroic properties of $RMnO_3$ oxides are primarily determined by the balance of ferromagnetic ($J_1$) and antiferromagnetic ($J_2$) in-plane interactions. Bulk YMO appears to be close to the bicritical point in the phase diagram [7] and thus a strong sensitivity of its magnetic order on small changes of cell parameters and bond angles could be expected.

In order to address the expected changes in the present set of samples with varying strain, it is worth to recall that the magnetic interactions in these manganites are mainly governed by the Mn-O-Mn bond angle ($\theta$). Available structural data in thin films do not allow accurate determination of $\theta$ as only (*a,b,c*) cell parameters are accessible and therefore quantification of the relative variation of in-plane ferromagnetic ($J_1$) and antiferromagnetic ($J_2$) interactions can not be done rigorous. In a previous paper where magnetic interactions in similar $YMnO_3$ thin films were discussed [14], we used $V_{uc}$ as a measure of the deformation of the YMO unit cell under strain. Here, taking into account that the most relevant parameter determining the magnetic interactions is $\theta$, we propose to use the ratio of the basal cell parameters (*a/b*) to rationalize the variation of $J_2/J_1$ under strain and subsequently the evolution of YMO films within the multiferroic phase diagram under strain. We notice that the variation of the *a/b* ratio under strain is identical to that of (*1/b*) as used in Ref. [21] as *a* is relaxed and constant for all films. The *a*/*b* ratio is thus a measure of the frustration of the magnetic interactions within the basal *ab* plane of the perovskite and thus it should determine the magnetic order, either collinear A-type, non-collinear (cycloidal) or collinear E-type. Indeed, A-type antiferromagnets have larger *a*/*b* values (0.963 in $LaMnO_3$ [22]) than cycloidal (0.904 in $TbMnO_3$ [23]) and than E-type (0.900 in $HoMnO_3$ [24]).

In Table I we included the measured cell parameters as well as the calculated *a*/*b* ratio. We note than more strained films display larger *a/b* values. This rule holds for films of both textures and in agreement to the above discussion is expected that any strain-induced magnetic canting should scale with a/*b* ratio. A clear correlation between these two parameters is observed (Fig. 9). For completeness, we also include the reported remanent magnetization for illustrative examples of E-type antiferromagnets ($HoMnO_3$, bulk YMO), cycloidal ($TbMnO_3$) and A-type ($LaMnO_3$ [25]) perovskites and the corresponding *a/b* ratio.

According to the structural trend observed in bulk materials described above, this observation would imply that under strain the structure of YMO films is pushed towards A-type structure, which is characterized by larger *a/b* ratios. No ferroelectricity has been reported nor predicted in A-type antiferromagnets and accordingly, in our films the signatures of cycloidal order and polarization switching disappear when increasing strain. Similarly, the peaks in the temperature dependence of the permittivity are washed out when reducing thickness and thus increasing strain.

Finally, it should be noticed that whereas switching is well visible in $\Delta\varepsilon_a(H)$, this is not the case of $\Delta\varepsilon_c(H)$ where its signature is rather subtle (Fig. 6(d-f) insets). This difference is rather surprising as it should be expected that if *c*-textured films have also the *bc*-cycloidal structure similar to the *a*-textured films, then the observed reduction of $P_c(H)$ [11] concomitant to the switching of the *bc*-cyloids towards *ac*-cycloids should produce similarly visible traces also in the $\Delta\ \varepsilon_c(H)$. This is not the case, as signatures of switching in $\Delta\varepsilon_c(H)$ are much weaker. At present we do not have a definitive explanation for this experimental observation. However we may tentatively propose that the existence of a mosaic of small 90° rotated domains in the *c*-textured films may lead either to the cancelation of *bc*-cycloids, much in the same way as reported in $BiFeO_3$ [26]. Alternatively, the small size of the domains in *c*-textured films may increase the switching field beyond the available range (<9 T).

From a more general perspective, it is striking to notice that bulk YMO being close to the bicritical point separating cycloidal and E-type regions, strain pushes the magnetic structure all the way across the phase diagram from the non-collinear, cycloidal region up to the A-type by increasing the *a/b* ratio. The observation (Fig. 9) of a gradually enhanced net magnetic moment is fully consistent with the stabilization of an A-type structure such that of $LaMnO_3$, where the occurrence of canting and net magnetic moment has been long ago documented [25]. We would point out that the disappearance of traces of 90º switchable polarization in films of increasing larger strain could, in principle, be attributed to the transformation of the cycloidal phase into E-type under strain. This alternative picture, is neither supported by the trend observed in the *a/b* ratio nor by the observation of a net magnetization in strained films, as in E-type structures any net magnetization arising from canting is cancelled due to symmetry reasons [27].

**Summary**

In summary, the dielectric properties of orthorhombic [001] and [100]-oriented $YMnO_3$ thin films have been explored as a function of film thickness and film texture. It has been observed that the thicker films display ferroelectric properties compatible with a cycloidal magnetic structure but not with an E-type one. In thinner and more in-plane compressively strained films, the ferroelectric polarization gradually disappears and a net magnetic moment emerges. It is argued that these observations are signatures of the progressive destabilization of the *bc*-cycloidal antiferromagnetic structure in benefit of a canted A-type magnetic structure.

**Acknowledgements**

Financial support by the Ministerio de Ciencia e Innovación of the Spanish Government [Projects MAT2008-06761-C03 and NANOSELECT CSD2007-00041] and Generalitat de Catalunya (2009 SGR 00376 ) is acknowledged.

References

[1] T.Kimura, et al., Nature **426**, 55 (2003).

[2] T. Goto, et al., Phys. Rev. Lett. **92**, 257201 (2004).

[3] H. Katsura, N. Nagaosa and A.V. Balatsky, Phys. Rev. Lett. **95**, 057205 (2005).

[4] M. Mostovoy Phys. Rev. Lett. **96**, 067601 (2006).

[5] I. A. Sergienko and E. Dagotto, Phys. Rev. B **73**, 094434 (2006).

[6] B. Lorenz, Y.-Q. Wang and Ch.-W. Chu, Phys. Rev. B **76**, 104405 (2007).

[7] S. Ishiwata, et al., Phys. Rev. B **81**, 100411(R) (2010).

[8] S. Picozzi, et al., Phys. Rev. Lett. **99**, 227201 (2007).

[9] K. Yamauchi, et al, Phys. Rev. B **78**, 014403 (2008).

[10] I. A. Sergienko, C. Sen and E.Dagotto, Phys. Rev. Lett. **97**, 227204 (2006).

[11] I. Fina et al., submitted

[12] I. C. Infante, F. Sanchez, J. Fontcuberta, M. Wojcik, E. Jedryka, S. Estrade, F. Peiro, J. Arbiol, V. Laukhin, and J. P. Espinos, Phys. Rev. B **76**, 224415 (2007).

[13] X. Marti, et al., Appl. Phys. Lett. **95**, 142903 (2009).

[14] X. Marti, et al., J. Appl. Phys. (in print).

[15] C J M Daumont, D Mannix, Sriram Venkatesan, G Catalan, D Rubi, B J Kooi, J Th M De Hosson and B Noheda, J. Phys.: Condens. Matter 21, 182001 (2009).

[16] A. Muñoz et al., J. Phys.: Condens. Matter **14**, 3285(2002).

[17] X. Marti, et al., Thin Solid Films **516**, 4899 (2008).

[18] J. F. Scott, et al., J. Appl. Phys. **64**, 787 (1988).

[19] I. Fina, et al., Thin Solid Films **518**, 4710 (2010).

[20] M. N. Iliev, M. V. Abrashev, H.-G. Lee, V. N. Popov, Y. Y. Sun, C. Thomsen, R. L. Meng, and C. W. Chu, Phys. Rev. B **57**, 2872 (1998).

[21] X. Marti, V. Skumryev, A. Cattoni, R. Bertacco, V. Laukhin, C. Ferrater, M.V. García-Cuenca, M. Varela, F. Sánchez, and J. Fontcuberta, Journal of Magnetism and Magnetic Materials **321**, 1719 (2009).

[22] F. Moussa, M. Hennion, J. Rodriguez-Carvajal, and H. Moudden, Phys. Rev. B **54**, 15149 (1996).

[23] M. Kenzelmann, A. B. Harris, S. Jonas, C. Broholm, J. Schefer, S. B. Kim, C. L. Zhang, S.-W. Cheong, O. P. Vajk, and J.W. Lynn, Phys. Rev. Lett. 95, 087206

(2005).
[24] A. Muñoz, M. T. Casáis, J. A. Alonso, M. J. Martínez-Lope, J. L. Martínez, and T. Fernández-Díaz, Inorg. Chem. 40, 1020 (2001).
[25] V. Skumryev, F. Ott, J.M.D. Coey, A. Anane, J.-P. Renard, L. Pinsard-Gaudart, and A. Revcolevschi, Eur. Phys. J. B **11**, 401 (1999).
[26] F. Bai, J. Wang, M. Wuttig, J.-F. Li, N. Wang, A. P. Pyatakov, A. K. Zvezdin, L. E. Cross, and D. Viehland, Appl. Phys. Lett. **86**, 032511 (2005).
[27] S. Picozzi and L. Chapon, during European School of Multiferroics 2010 (L'Aquila, Italy; October 2010). Private communication.

**Figure Captions**

**Fig 1.** Diffracted X-ray intensity distribution at the scattering angle corresponding to the YMO(111) plane collected at a wide range of azimuthal and polar angles in films grown on (a) STO(110) and (e) STO(001) substrates. The films are epitaxial and [100] and [001] textured, respectively. Panels (b-d) show the reciprocal space maps around the (311) reflections of the (100) textured films. Panels (f-h) show reciprocal space maps showing (208) and (028) reflections of the [001] textured samples. Films thicknesses are indicated.

**Fig 2.** Dependence of dielectric permittivity on temperature. Sketches in (a,d) and axis in (b,e) indicate the sample orientation for dielectric measurements. The existence of two domains 90º rotated in c-textured films is indicated. (c,f) Temperature dependence of the dielectric permittivity along a-axis and c-axis respectively for films of various thicknesses as indicated.

**Fig 3.** Electric-field dependent dielectric permittivity (left axis) and losses (tg $\delta$) (right axis) measured at 5 K for *a*-textured $\varepsilon_a$(E) (a, b, c) and *c*-textured $\varepsilon_a$ (d, e, f) films of indicated distinct thickness.

**Fig 4.** Temperature dependence of the integrated remanent charge (Q) extracted from field-dependent cycles of the 106 nm *a*-textured (squares) and 137 nm *c*-textured (circles) films.

**Fig 5.** Current versus voltage loops measured at 5 K using PUND for *a*-oriented samples at 10 kHz under a magnetic field (6 T // *b*) (a, b, c), and for *c*-oriented samples at 1kHz (d, e, f) (H=0).

**Fig 6.** Dielectric permittivity $\Delta\varepsilon(H)$ (= $[\varepsilon(H)-\varepsilon(H=0)]/\varepsilon(H=0)$) along *a*-axis *vs* magnetic field applied along *a*, *b* and *c*-axis for films of a) 106 nm, b) 50 nm and c) 24 nm. In (d, e and f) the dielectric permittivity along c-axis *vs* magnetic field applied along *ab*-plane and *c*-axis for films of 137, 64 and 28 nm, respectively. Expanded views of the field region

where switching occurs are zoomed (insets in (d,e, f)).

**Fig 7.** Dielectric permittivity $\Delta\varepsilon(H)$ (= $[\varepsilon(H)-\varepsilon(H=0)]/\varepsilon(H=0)$) along: a) *a*-axis and b) *c*-axis *vs* $H^2$. The magnetic field has been applied along *b*-axis and along the *ab*-plane as indicated.

**Fig 8.** a) Temperature dependence of the magnetization after ZFC and FC processes of *a*-textured films of various thicknesses. Measurements have been done using a field of 50 mT applied parallel to *a*-axis. b) Field dependent magnetization measured at 5 K with the magnetic field applied along the *c*-axis of the same films of a).

**Fig 9.** Remanent magnetization measured at 5K ($M_r$) (open symbols) and magnetization after Field-Cooling process (50 mT, 25 K) ($M_{FC}$) (solid and partially filled symbols) of the *a*-textured (circles) and *c*-textured films (triangles). Data for *c*-textured films from Ref. 13 and 14 are indicated by open and partially filled triangles. $M_{FC}$ data for $HoMnO_3$ [24] (symbols), bulk YMO [20] (symbols), ($TbMnO_3$ [23] (symbols) and A-type ($LaMnO_3$ [22, 25]) (symbols) are included.

**Table caption**

**Table I.** Cell parameters extracted from reciprocal spaces maps of *a*- and *c*- textured thin films of various thicknesses.

**Figure 1**

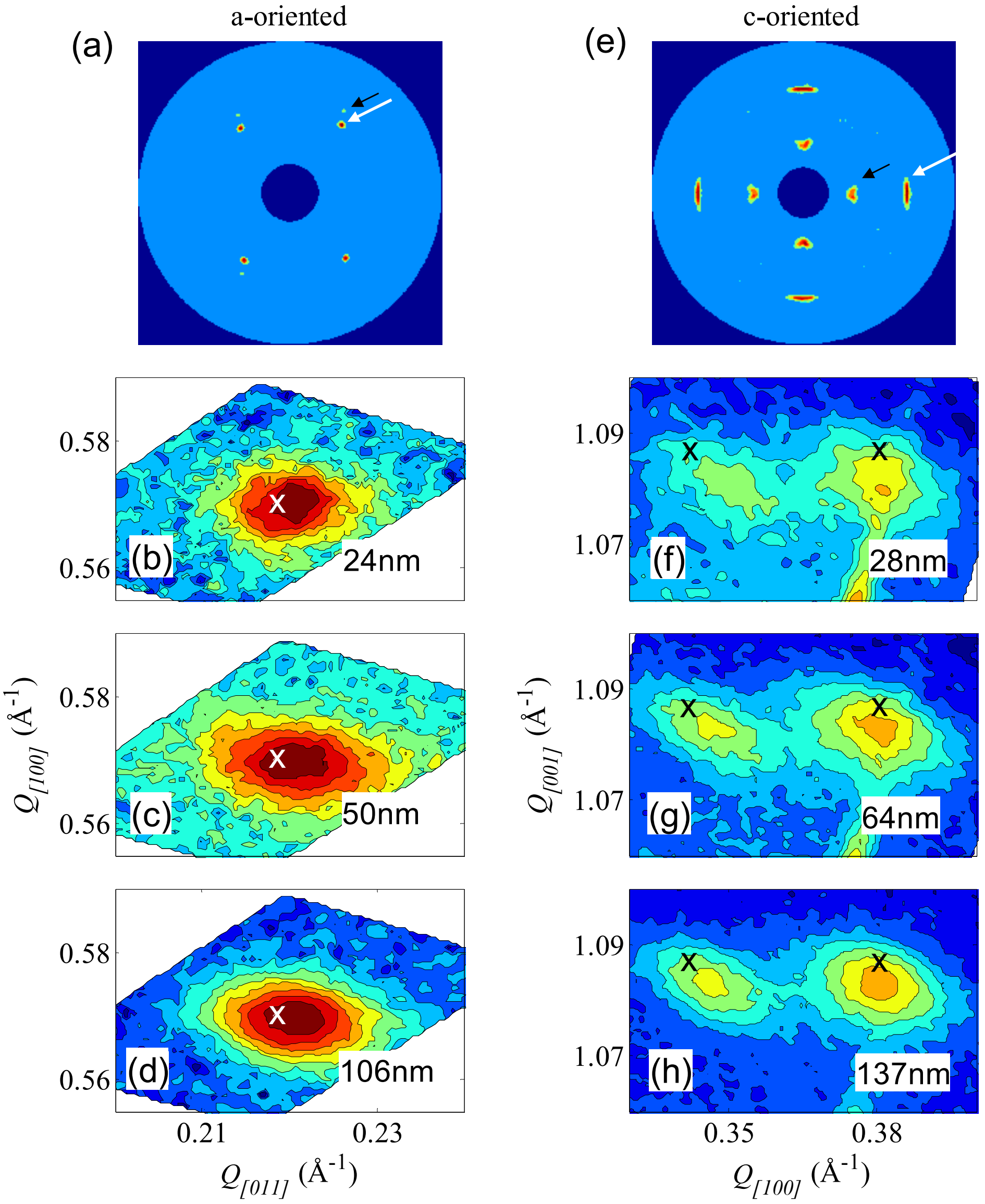

**Figure 2**

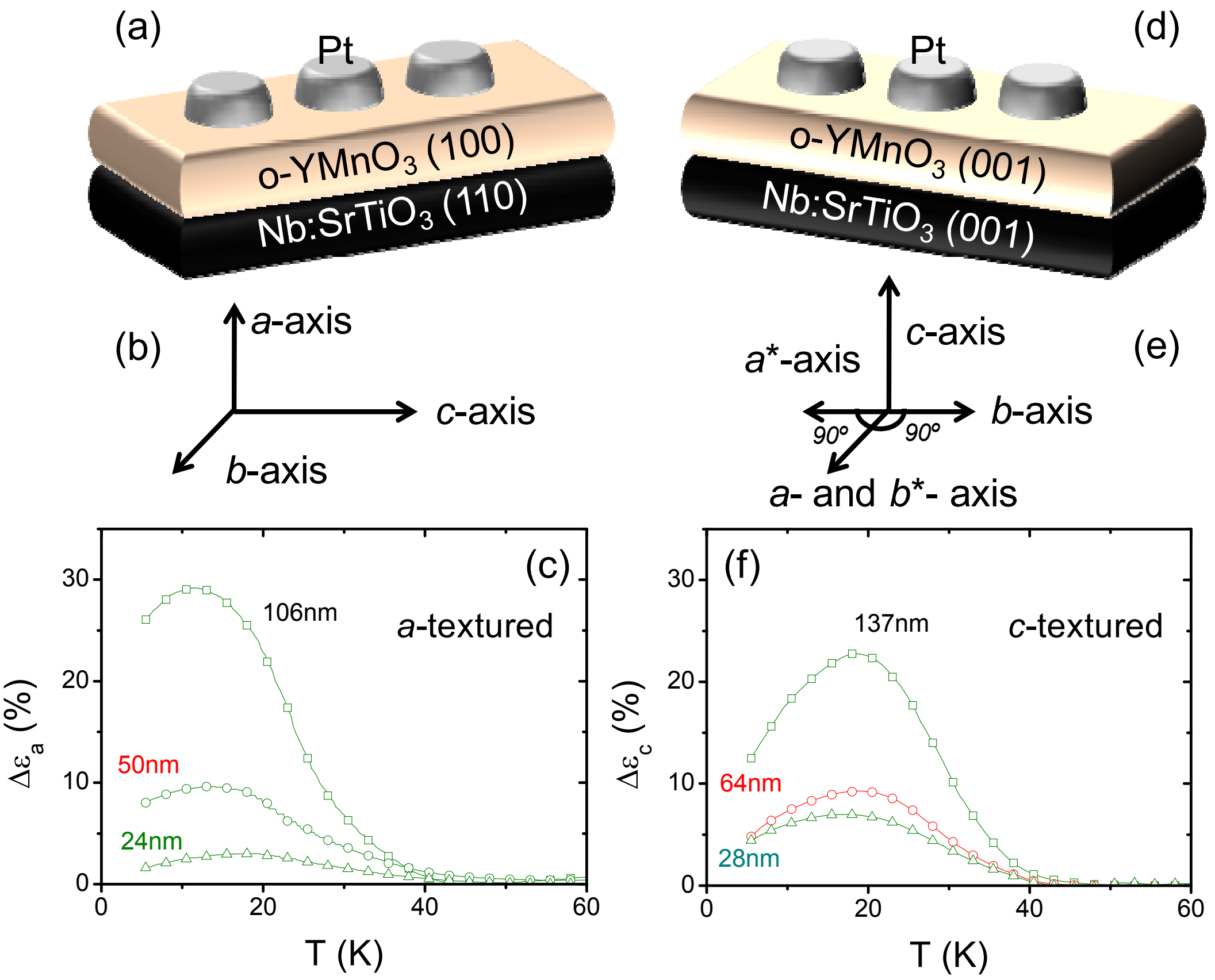

**Figure 3**

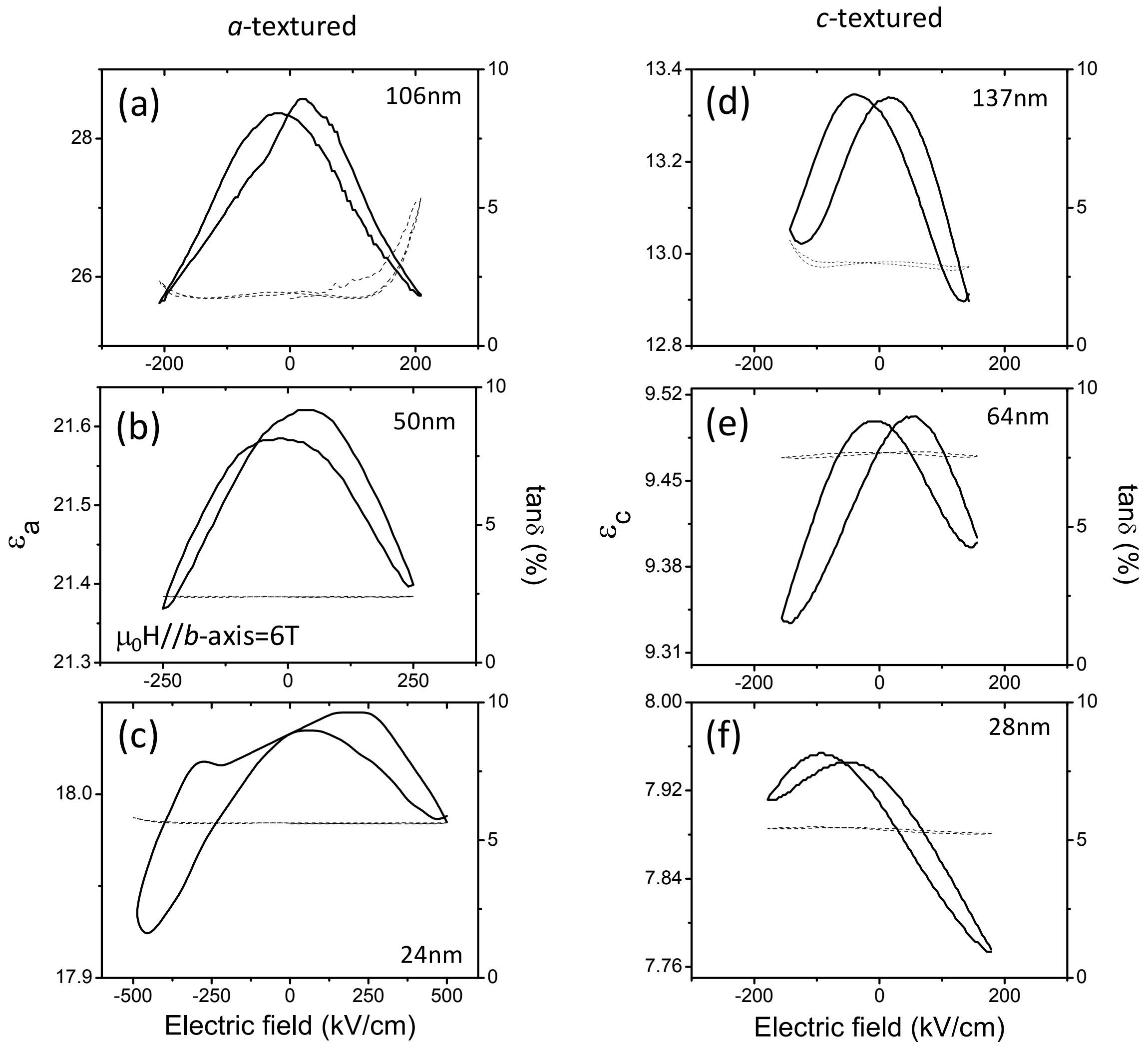

**Figure 4**

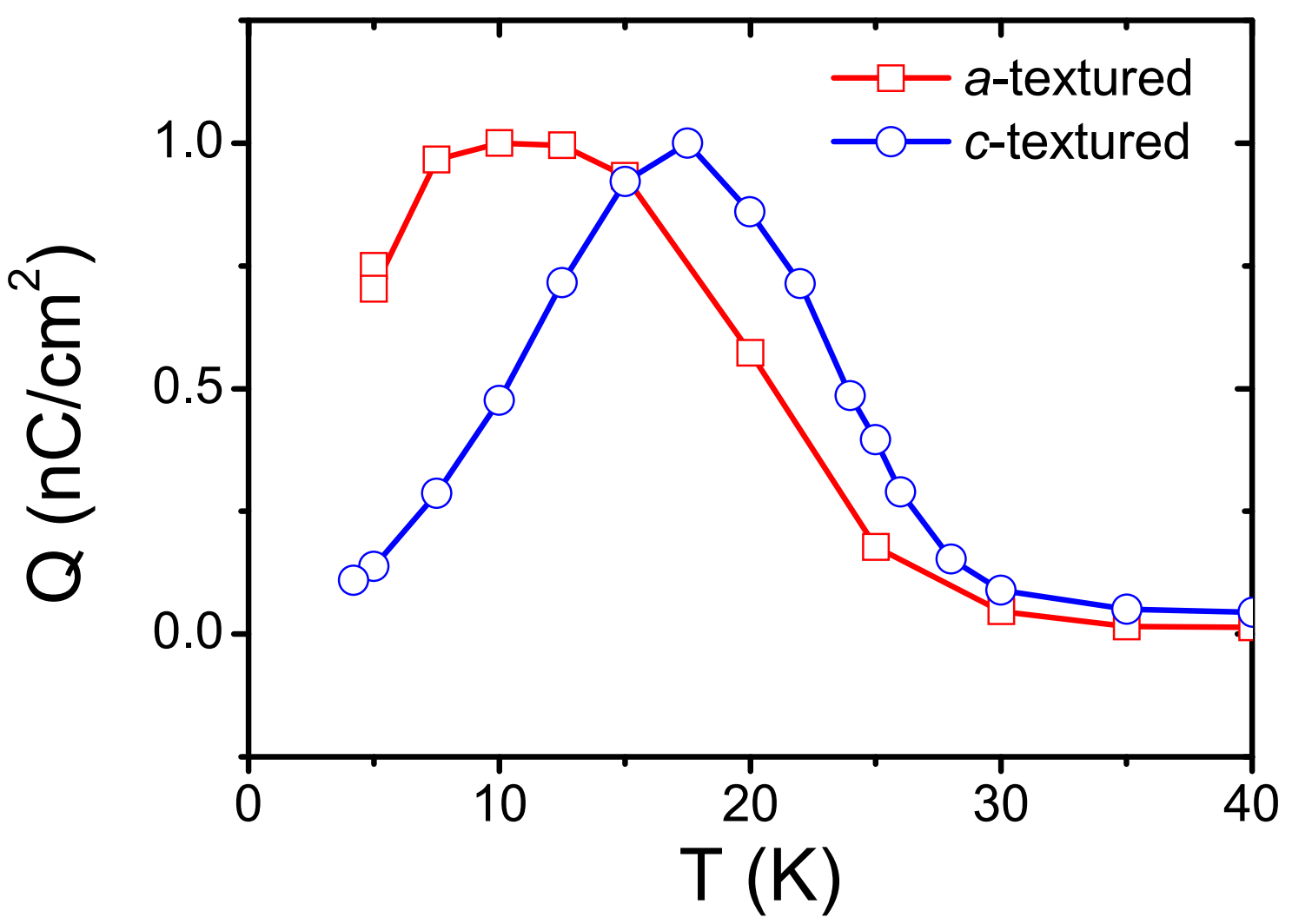

**Figure 5**

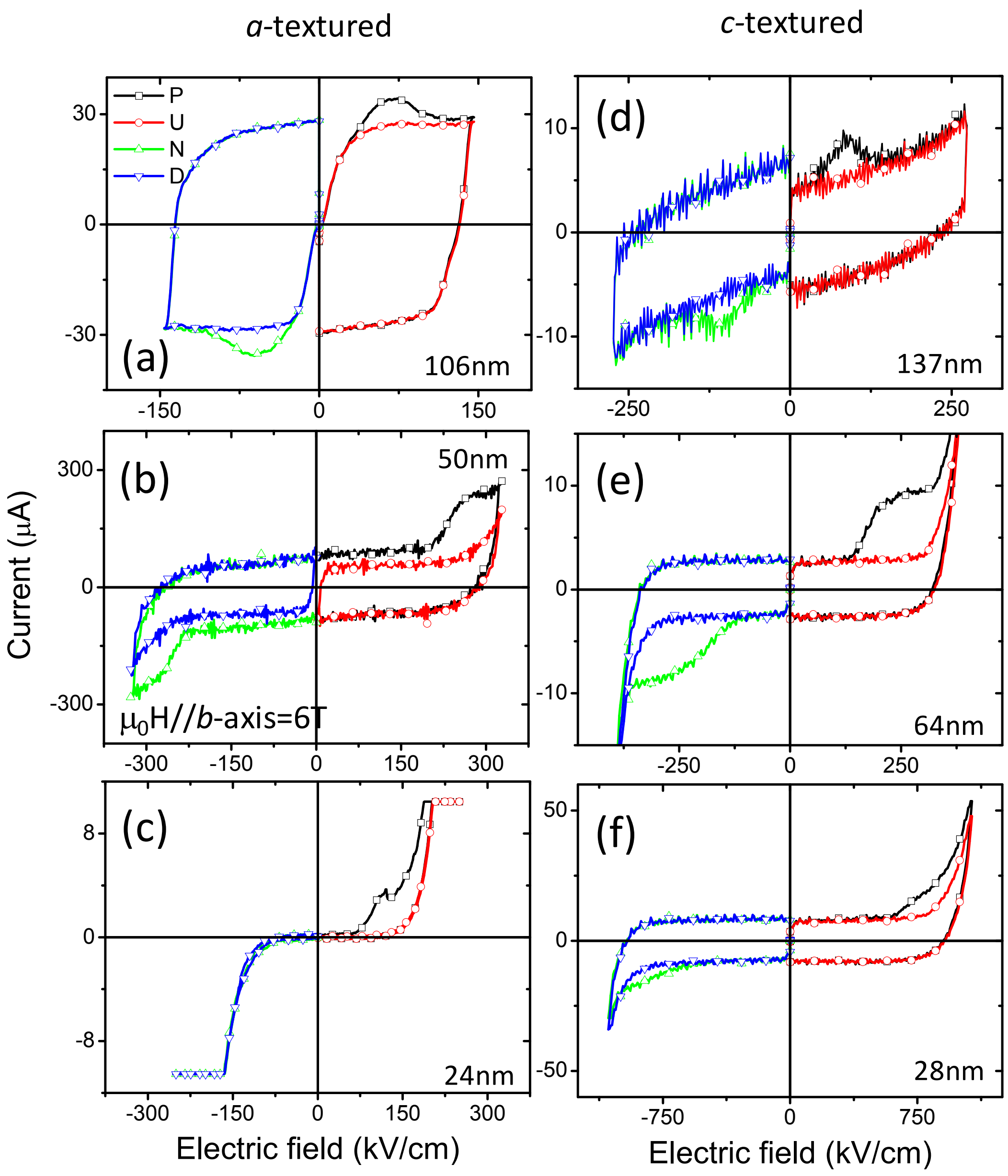

**Figure 6**

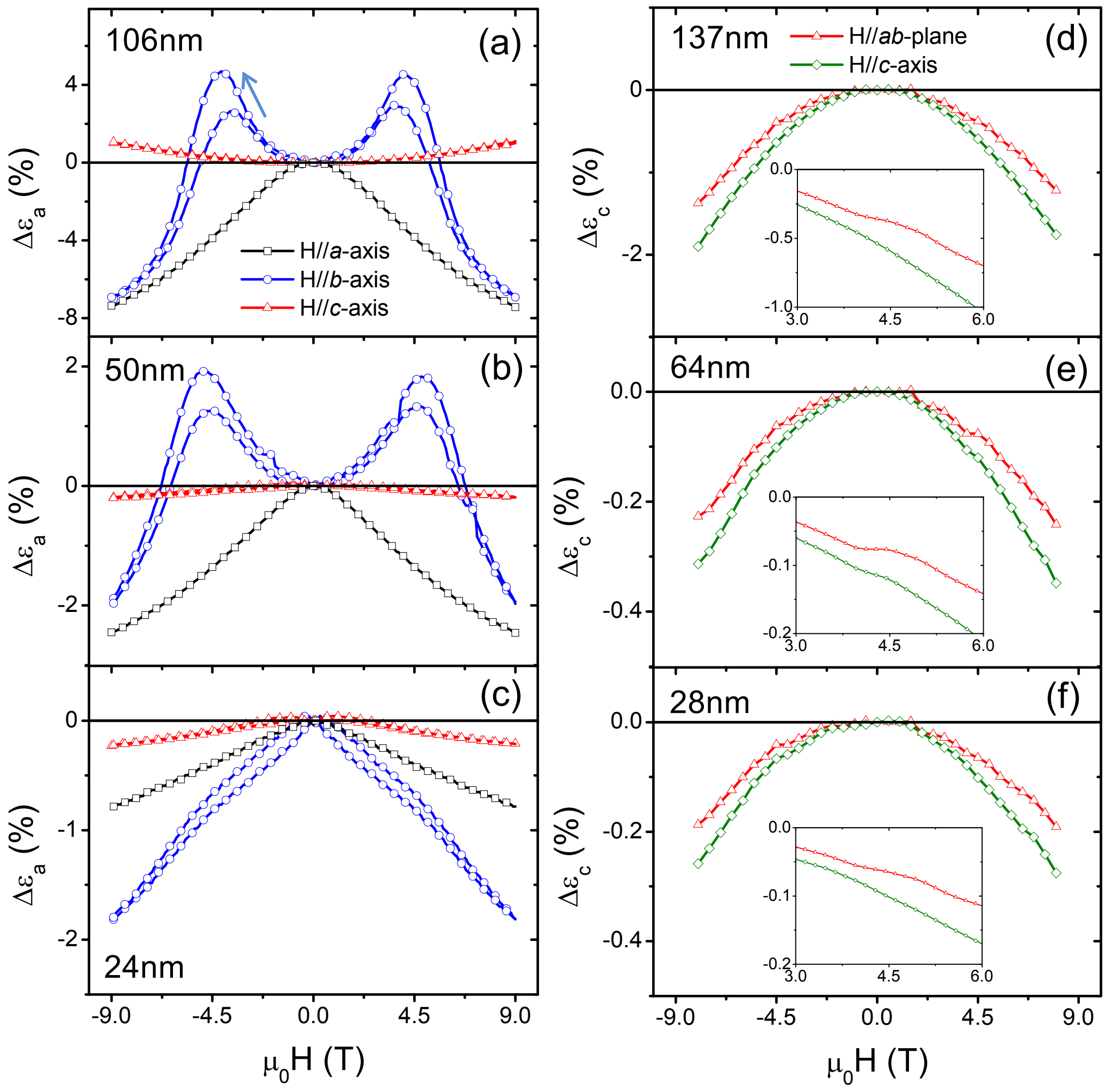

**Figure 7**

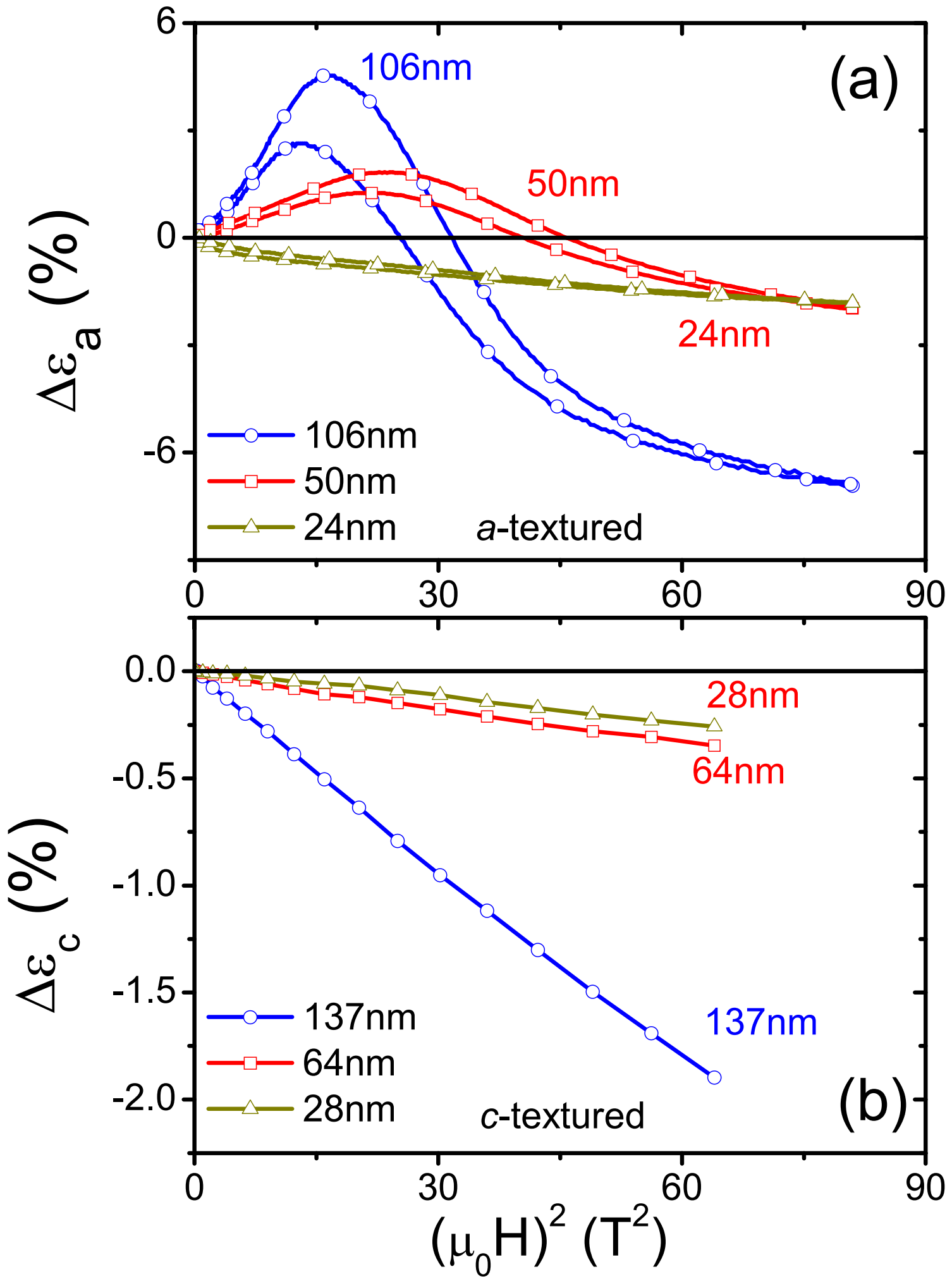

**Figure 8**

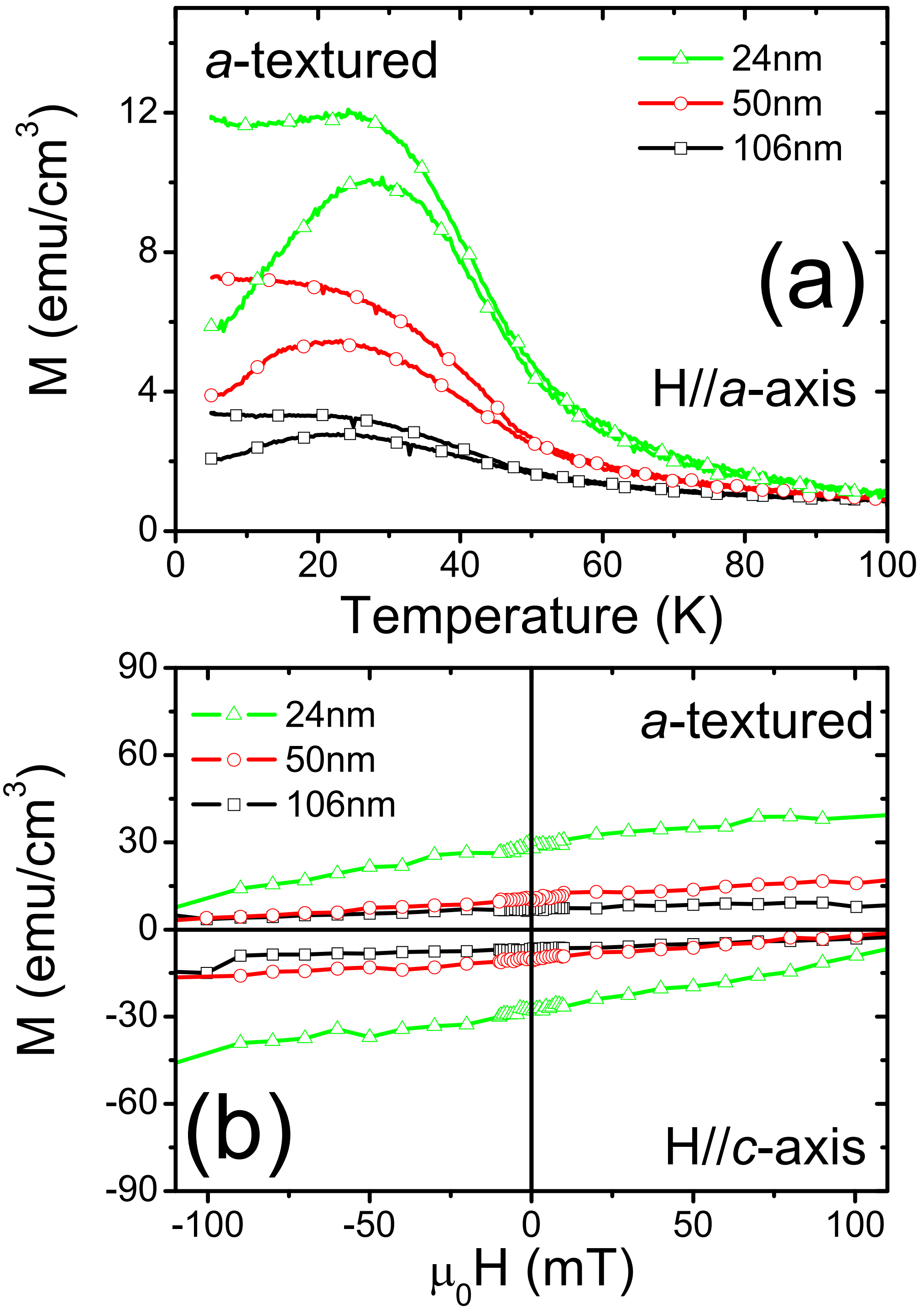

**Figure 9**

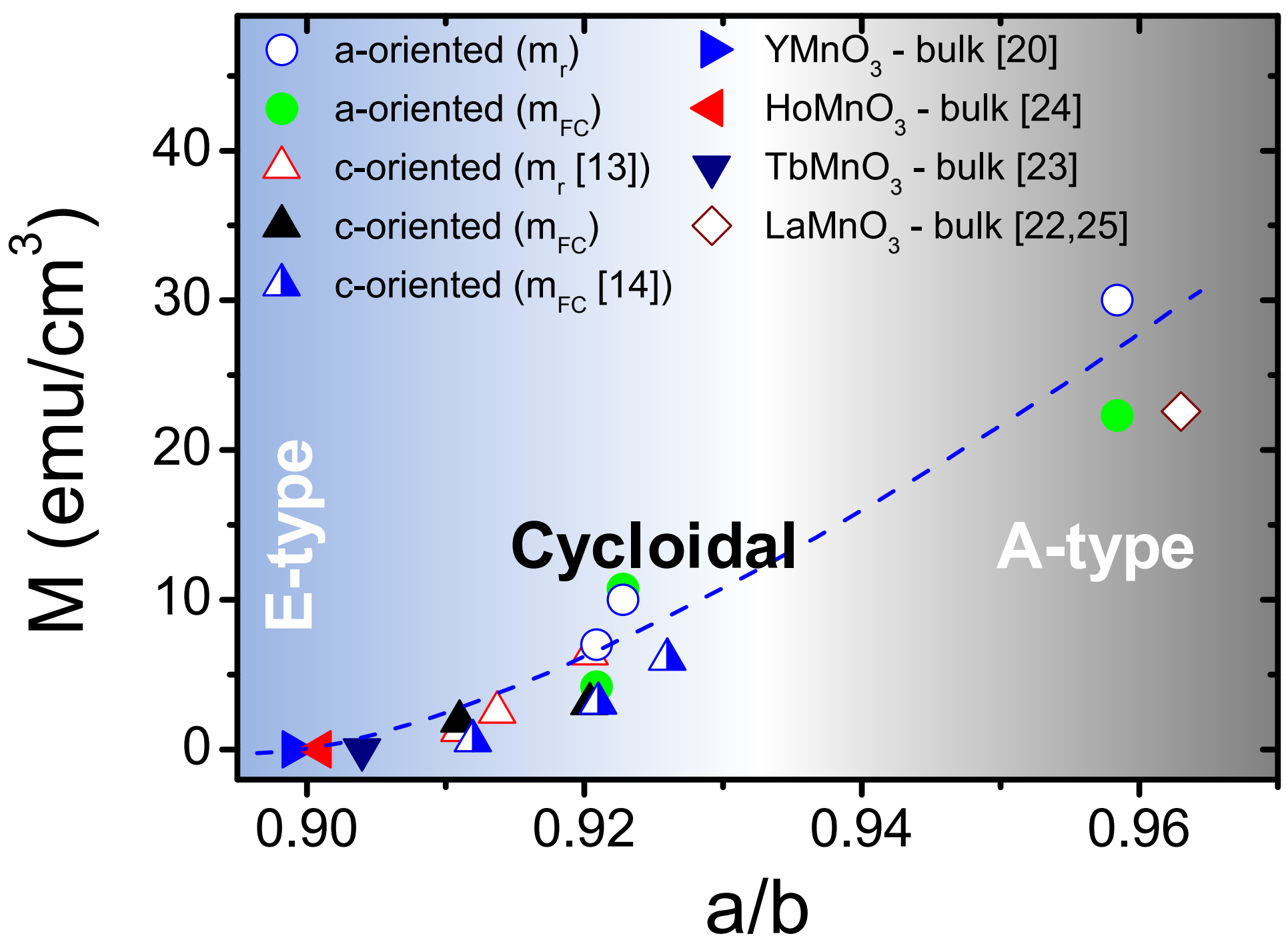

| **Table I** | | | | | |
|---|---|---|---|---|---|
| ***a*-textured** | *a* (Å) | *b* (Å) | *c* (Å) | $V_{uc}$ (Å$^3$) | *a/b* |
| 106 nm | 5.260 | 5.712 | 7.453 | 223.93 | 0.921 |
| 50 nm | 5.260 | 5.700 | 7.430 | 222.77 | 0.923 |
| 24 nm | 5.257 | 5.485 | 7.837 | 225.98 | 0.958 |
| | | | | | |
| ***c*-textured** | | | | | |
| 137 nm | 5.260 | 5.774 | 7.377 | 224.03 | 0.911 |
| 64 nm | 5.260 | 5.757 | 7.393 | 223.86 | 0.914 |
| 28 nm | 5.260 | 5.715 | 7.398 | 222.38 | 0.920 |
| | | | | | |
| Bulk [20] | 5.26 | 5.85 | 7.36 | 226.4 | 0.899 |